\documentclass[aps,prl,10pt,twocolumn,superscriptaddress]{revtex4-2}

\usepackage{times}
\usepackage{changes}
\usepackage{graphicx}
\usepackage{amssymb}
\usepackage{epstopdf}
\usepackage{amsmath}
\usepackage{color}
\usepackage{hyperref}
\usepackage{bbold}
\usepackage{bm}
\usepackage{enumitem}
\usepackage[acronym]{glossaries}
\usepackage{physics}
\makeatletter
\graphicspath{{Images/}}

\usepackage{hyperref}
\hypersetup{
	colorlinks=true,
	linkcolor=black,          
	citecolor=blue,           
	filecolor=magenta,        
	urlcolor=cyan
}

\let\oldbm\bm
\renewcommand{\bm}[1]{\oldbm{\mathrm{#1}}} 

\newcommand{\hsig}{\hat{\bm{\sigma}}} 
\newcommand{\hs}{\hat{\bm{s}}} 

\makeatother	

\begin{document}
	
	\title{Genuine quantum scars in many-body spin systems}
	
	\author{Andrea Pizzi}
	\email{ap2076@cam.ac.uk}
	\affiliation{Cavendish Laboratory, University of Cambridge, Cambridge CB3 0HE, United Kingdom}
	\affiliation{Department of Physics, Harvard University, Cambridge, Massachusetts 02138, USA}
	
	\author{Long-Hei Kwan}
	\affiliation{Cavendish Laboratory, University of Cambridge, Cambridge CB3 0HE, United Kingdom}
	
	\author{Bertrand Evrard}
	\affiliation{Laboratoire Matériaux et Phénomènes Quantiques, Université Paris Cité, 75013 Paris, France}
	
	\author{Ceren B.~Dag}
	\affiliation{Department of Physics, Indiana University, Bloomington, Indiana 47405, USA}
	\affiliation{Department of Physics, Harvard University, Cambridge, Massachusetts 02138, USA}
	\affiliation{ITAMP, Center for Astrophysics, Harvard $\&$ Smithsonian, Cambridge, Massachusetts 02138, USA}
	
	\author{Johannes Knolle}
	\email{j.knolle@tum.de}
	\affiliation{Department of Physics, Technische Universit{\"a}t M{\"u}nchen TQM, James-Franck-Stra{\ss}e 1, 85748 Garching, Germany}
	\affiliation{Munich Center for Quantum Science and Technology (MCQST), 80799 Munich, Germany}
	\affiliation{Blackett Laboratory, Imperial College London, London SW7 2AZ, United Kingdom}
	
	\begin{abstract}
		Chaos makes isolated systems of many interacting particles quickly thermalize and forget about their past. Here, we show that quantum mechanics hinders chaos in many-body systems: although the quantum eigenstates are thermal and strongly entangled, exponentially many of them are scarred, that is, have an enlarged weight along underlying classical unstable periodic orbits. Scarring makes the system more likely to be found on an orbit it was initialized on, retaining a memory of its past and thus weakly breaking ergodicity, even at long times and despite the system being fully thermal and the eigenstate thermalization hypothesis fulfilled. We demonstrate the ubiquity of quantum scarring in many-body systems by considering a large family of spin models, including some of the most popular ones from condensed matter physics. Our findings, at hand for modern quantum simulators, prove structure in spite of chaos in many-body quantum systems.
	\end{abstract}
	
	\maketitle
	
	\section{Introduction}
	Understanding and controlling many-body quantum systems out of equilibrium is a key challenge of modern physics. Left to their isolated dynamics, as becoming increasingly possible in ever improving quantum computers and simulators~\cite{gross2017quantum, blatt2012quantum, zhang2017observation, bernien2017probing, monroe2021programmable, mi2022time}, these systems tend to quickly relax to thermal equilibrium, effectively forgetting about their past in the spirit of ergodicity~\cite{polkovnikov2011colloquium,eisert2015quantum,d2016quantum}. This fate is as universal as tame, and underpinning it is the seemingly chaotic nature of the many-body Hamiltonian. Indeed, the statistical properties of the system's eigenvalues and eigenvectors are in many respect similar to those of certain random matrices~\cite{atas2013distribution}.
	
	It is known for single-particle systems, however, that chaos can be hindered by quantum scarring~\cite{heller1984bound, berry1989quantum, kaplan1999scars}. This is the phenomenon whereby the quantum wavefunction is enhanced along underlying classical unstable periodic orbits (UPOs). Scarred eigenstates are less random than chaos would suggest, increasing the chances of finding the system on a UPO it was prepared on and challenging the notion of ergodicity~\cite{kaplan1998linear, kaplan1999scars}. Scarring has long been known within single-particle quantum chaos, but its generalization to many-body quantum systems, whose study requires modern quantum simulators and more advanced numerical tools, has remained virtually unexplored, limited to specific recent instances of interacting bosons in a ring lattice~\cite{hummel2023genuine} and a periodically driven spin-1 chain~\cite{evrard2024quantumb}.
	
	Here, we show that scarring is ubiquitous in many-body systems. For a large family of spin chains, we find that exponentially many eigenstates are enhanced along the UPOs of the associated classical dynamics. Initializing the system on a UPO enhances the probability of finding it on the same UPO at later times. Even in the middle of the spectrum, where entanglement is close to maximal and the eigenstate thermalization hypotesis (ETH) fulfilled~\cite{rigol2008thermalization}, we show that the eigenstates are less chaotic than expected and ergodicity is weakly broken.
	
	Our work adds quantum scarring to the (short) list of mechanisms yielding nontrivial effects in many-body quantum systems out of equilibrium, such as integrability~\cite{rigol2007relaxation, evrard2024quantuma}, many-body localization~\cite{pal2010many, alet2018many, abanin2019colloquium, sels2021dynamical}, Hilbert space fragmentation~\cite{sala2020ergodicity, khemani2020localization}, and non-thermal eigenstates in an otherwise chaotic spectrum~\cite{moudgalya2018entanglement, turner2018weak, ho2019periodic, serbyn2021quantum, moudgalya2022quantum}. All these rely on an explicit or emergent partial integrability and host ETH-breaking eigenstates. By contrast, scarring establishes a deviation from chaos in the thermal eigenstates of generic non-integrable many-body systems, where one would least expect it.
	
	A remark on nomenclature is due. Growing attention has been recently devoted to certain many-body quantum systems hosting a few eigenstates which violate ETH and are weakly entangled~\cite{turner2018weak, ho2019periodic, serbyn2021quantum, moudgalya2022quantum}. While these have been dubbed ``quantum many-body scars'', there is no evidence that they are in fact scars, because they could not be related to UPOs in a chaotic phase space, but rather were often associated to its regular regions~\cite{ho2019periodic,michailidis2020slow,turner2021correspondence,lerose2023theory,omiya2024quantum,muller2024semiclassical,kerschbaumer2024quantum}.
	Such athermal eigenstates are not the focus of our work. Instead, here we consider the thermal eigenstates of many-body systems and show their genuine scarring, due to UPOs in a chaotic phase space.
	
	Beyond~\cite{hummel2023genuine,evrard2024quantumb}, note that such genuine scarring has also been shown in the Dicke model~\cite{pilatowsky2021ubiquitous} and for a spinor condensate~\cite{evrard2024quantuma} (also observed in experiments~\cite{austin2024observation}), that with their all-to-all interactions sit somewhere in between few- and many-body systems. Moreover, a semiclassical analysis of the periodic orbits of a quantum many-body system was presented for a periodically driven spin chain in~\cite{akila2017semiclassical}, although not in relation to scarring.
	
	\section{Results}
	
	\textit{Model, classical dynamics, and UPOs} ---
	Consider a chain of $N$ spin $s$ particles subject to a magnetic field and nearest-neighbor interactions ($\hbar = 1$)
	\begin{equation}
		\hat H
		= \sum_{j = 1}^N \left( \bm{\mu} \cdot \hs_j + \frac{1}{s} \hs_j \bm{J} \hs_{j+1} \right),
		\label{eq. H}
	\end{equation}
	where $\hs_j^2 = s(s+1)$, $\hs_j \bm{J} \hs_{j+1} = \sum_{\alpha,\beta = x,y,z} \hat{s}^{\alpha}_j J_{\alpha\beta} \hat{s}^{\beta}_{j+1}$, $J_{\alpha\beta} = J_{\beta\alpha}$, $\bm{\mu} \cdot \hs_j = \sum_{\alpha = x,y,z} \mu_{\alpha} \hat{s}^{\alpha}_j$, and periodic boundary conditions are assumed. This Hamiltonian is very general, e.g., it describes \textit{any} homogeneous spin $1/2$ chain with nearest-neighbor reciprocal interactions, including many prototypical models of condensed matter physics. For instance, the Ising model with both transverse and longitudinal (integrability-breaking) fields is obtained for $\mu_y = 0$ and $\bm{J} = J_{zz} \bm{z} \otimes \bm{z}$, yielding
	$\hat H = \frac{1}{2} \sum_{j} \left( \mu_x \hat{\sigma}^x_j + \mu_z \hat{\sigma}^z_j + J_{zz} \hat{\sigma}^z_j \hat{\sigma}^z_{j+1} \right)$,
	with $\hat{\sigma}^{x,z}_j$ standard Pauli operators. The Heisenberg, XX, and XXZ models are obtained in a similar way. The normalization $s^{-1}$ in front of the interaction in Eq.~\eqref{eq. H} ensures a well-defined limit $s \to \infty$, for which the spins can be described as classical rotors $\{ \bm{s}_j \}$, with $|\bm{s}_j|^2 = 1$ and dynamics~\cite{landau1935theory}
	\begin{equation}
		\frac{d \bm{s}_j}{dt} = 
		\left[\bm{\mu} + \bm{J} \left( \bm{s}_{j-1} + \bm{s}_{j+1} \right) \right]
		\cross \bm{s}_j.
		\label{eq. classical dynamics}
	\end{equation}
	
\begin{figure}
		\centering
		\includegraphics[width=\linewidth]{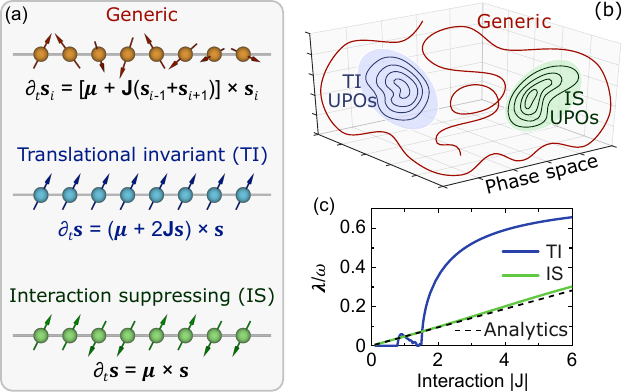}
		\caption{\textbf{Classical spin dynamics and unstable periodic orbits.}
			(a) The dynamics of a classical spin chain is generally chaotic, but yields unstable periodic orbits (UPOs) for the TI states, with aligned spins (blue), and for the IS states, with spins alternating every other site (green).
			(b) The TI states and IS states form two-dimensional manifolds of UPOs within the highly-dimensional phase space.
			(c) Quantum scarring is favoured for $\lambda/\omega < 1$, with $\lambda$ the Lyapunov exponent of the UPO and $\omega$ its frequency. This condition holds throughout the whole considered parameter regime, and especially for small $|\bm{J}|/|\bm{\mu}|$. In (c) we considered the Ising model with $\bm{\mu} = (2.4,0,0.4)$, $N = 100$, and the UPOs through $\bm{s} = \bm{y}$. The Lyapunov exponent is computed numerically from the monodromy matrix~\cite{contopoulos2002order}, and analytically (dashed line) for the IS states and $|\bm{J}| \ll |\bm{\mu}|$~\cite{SM}.
		}
		\label{Fig1}
	\end{figure}

	The nonlinear dynamics in Eq.~\eqref{eq. classical dynamics} is generally chaotic and aperiodic~\cite{de2012largest}. There are however special families of spin configurations for which the dynamics is periodic instead, see Fig.~\ref{Fig1}(a). One is that of the translationally invariant (TI) states, in which all the spins are aligned, $\{\bm{s}_j \} = \left( \bm{s},\bm{s},\bm{s},\bm{s},\dots\right)$. Another, for $N$ multiple of $4$, is that of the interaction suppressing (IS) states, in which the spins flip at every other site, $\{\bm{s}_j\} = \left(+\bm{s},+\bm{s},-\bm{s},-\bm{s},+\bm{s},+\bm{s},-\bm{s},-\bm{s}, \dots \right)$. These states are special in that the classical dynamics in Eq.~\eqref{eq. classical dynamics} does not destroy their nature: a TI state remains such, owing to translational invariance, and a IS state remains such, because $\bm{s}_{j-1} + \bm{s}_{j+1} = 0$ and the interaction is suppressed. The dynamics from these states is fully specified by the dynamics of just one spin, say $\bm{s}$, namely
	\begin{equation}
		\frac{d \bm{s}}{dt}
		=
		\begin{cases}
			\left(\bm{\mu} + 2 \bm{J} \bm{s} \right) \cross \bm{s}
			& \text{for TI states} \\
			\bm{\mu} \cross \bm{s}
			& \text{for IS states}
		\end{cases}.
		\label{eq. UPOs}
	\end{equation}
	The vector $\bm{s}$ lives on the surface of a sphere, and its Hamiltonian dynamics must be periodic, like any in two dimensions~\cite{strogatz2018nonlinear}.
	
	The many-body dynamics from a TI or IS state will also be periodic but, crucially, generally unstable: a slight perturbation that breaks the TI or IS character of the initial condition leads through chaos to a highly unpredictable and nonperiodic trajectory exploring the many-body phase space ergodically. For the IS states and $|\bm{J}| \ll |\bm{\mu}|$, we compute all the Lyapunov exponents \textit{analytically}~\cite{SM}, showing that for $N > 4$ the IS states are indeed unstable for all models but trivial ones (e.g., $\lambda = 0$ for the Ising model in a longitudinal field). The TI and IS states thus constitute two manifolds of UPOs within the classical many-body phase space, Fig.~\ref{Fig1}(b). The existence of continuous manifolds of UPOs, rather than isolated UPOs, is a favourable factor for scarring~\cite{kaplan1998linear}.
	
	We note that other such manifolds exist. the IS states are in fact part of a broader manifold of periodic orbits, namely $\{\bm{s}_j\}= \left(\bm{s}_1,\bm{s}_2,-\bm{s}_1,-\bm{s}_2,\bm{s}_1,\bm{s}_2,-\bm{s}_1,-\bm{s}_2,\dots \right)$, also yielding $\frac{d \bm{s}_i}{dt} = \bm{\mu} \cross \bm{s}_i$. We also note that the condition $N = 4k$ led to special effects also in~\cite{akila2017semiclassical}, resulting in periodic orbits with enhanced impact on the spectral properties of a periodically driven spin chain, and in~\cite{hummel2023genuine}, relating to ``hopping suppressing'' UPOs of bosons in a lattice.
	
	\begin{figure*}
		\centering
		\includegraphics[width=\linewidth]{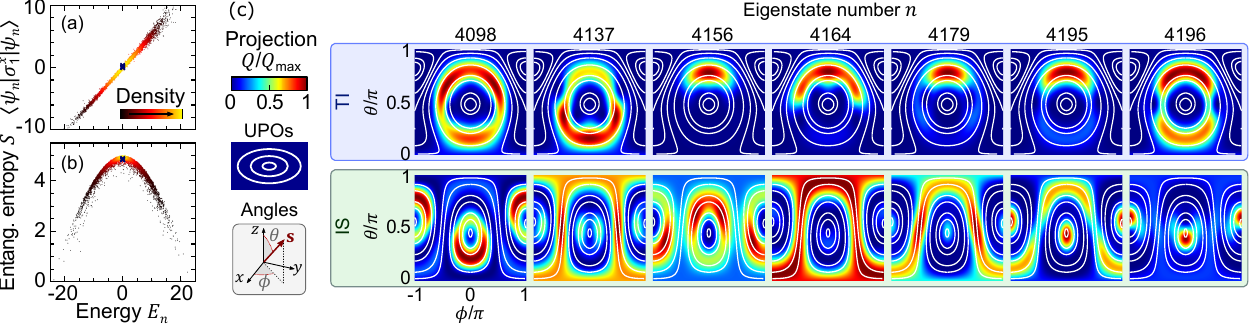}
		\caption{\textbf{Quantum scars in many-body spin chains.}
			(a,b) The many-body eigenstates $\ket{E_n}$ are fully thermal:
			the expectation value of local observables, e.g., $\bra{E_n} \hat{\sigma}^x_1 \ket{E_n}$, only depends on the energy~\cite{rigol2008thermalization}, as does the bipartite entanglement entropy $S$, which is extensive, $S \sim N$.
			(c) Projection of selected eigenstates $\ket{E_n}$ onto the TI and IS manifolds of the classical phase space (top and bottom, respectively). The many-body eigenstates are scarred, that is, enhanced along certain UPOs (white lines). The considered eigenstates are marked in (a,b) by crosses, and sit in the middle of the thermal spectrum. For each plot we define $Q_{\rm max} = \max_{\theta, \phi} Q(\theta,\phi)$. Here, we considered the Ising model with $s = \frac{1}{2}$, $\bm{\mu} = (2.4,0,0.4)$, $J_{zz} = -1.8$, and $N = 16$.}
		\label{Fig2}
	\end{figure*}
	
	\begin{figure}
		\centering
		\includegraphics[width=\linewidth]{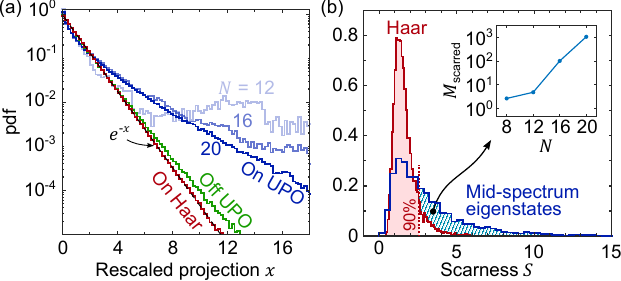}
		\caption{\textbf{Quantitative features of scarring.}
			(a) Distribution of the rescaled overlaps $x$ between mid-spectrum eigenstates and various families of states. The overlaps with Haar random states (red) have distribution $e^{-x}$ (dashed), similarly to the overlaps with generic points $\{\bm{s}_i\}$ of the phase space (green). By striking contrast, the overlaps with the IS states (blue) yield a much fatter tail, that is, scarring makes some eigenstates anomalously large on the UPOs. (b) The distribution of the scarness parameter $S$ for mid-spectrum eigenstates (blue) has a long tail compared to Haar random states (red). Indeed, the number $M_{\rm scarred}$ of scarred eigenstates, defined as the number of eigenstates with $S$ larger than the 90-th percentile of the Haar states (red dashed line), minus $10\%$~\cite{SM}, is large and appears to grow exponentially with system size (inset). In (a) and (b) the eigenstates are uniformly sampled from the middle of the spectrum, namely among the $10\%$ eigenstates with lowest $|E|$. In (a), the phase-space states are obtained sampling the spins $\{\bm{s}_i\}$ uniformly and independently from the sphere, and the IS states are obtained sampling $\bm{s}_1$ uniformly from the sphere and alternating the other spins at every other site as in Fig.~\ref{Fig1}(a). Here, $N = 20$ except where otherwise and explicitly specified. All other parameters are as in Fig.~\ref{Fig2}.}
		\label{Fig3}
	\end{figure}
	
	\begin{figure}
		\centering
		\includegraphics[width=\linewidth]{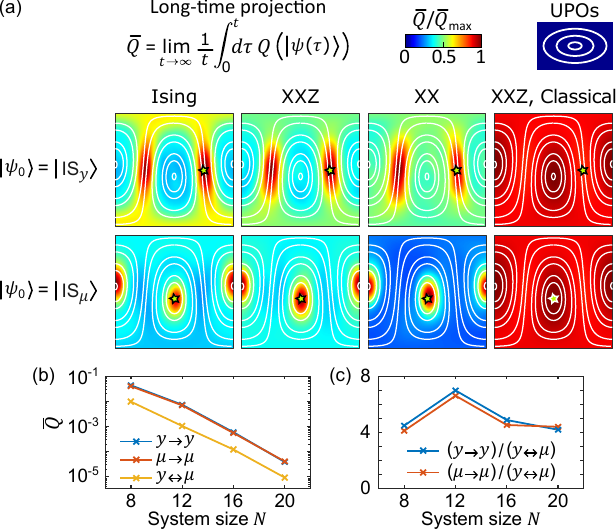}
		\caption{\textbf{Weak ergodicity breaking from quantum scarring.}
			(a) Time-averaged projection $Q(\ket{\psi(t)})$ over the manifold of IS states, for various models and initial conditions. In the top row, the system is initialized in $\ket{{\rm IS}_y}$, namely, on the IS UPO aligned along $\bm{y}$ ($\phi = \theta = \pi/2$, marked by a star), and is more likely to be found on the same UPO at long times. In the bottom row, the system is initialized in $\ket{{\rm IS}_\mu}$, namely, on the IS UPO aligned along $\bm{\mu}$ (also marked by a star), and is more likely to be found there at long times. That is, the system retains some information on its initial condition and weakly breaks ergodicity. This is a quantum effect: the classical projection~\cite{SM}, shown for the XXZ model, is insensitive to the initial condition and almost uniform.
			(b,c) Scaling of $\bar{Q}$ with system size $N$, with focus on the time-averaged return probabilities ($\bar{Q}(\bm{y})$ for $\ket{\psi_0} = \ket{{\rm IS}_y}$, denoted $y \rightarrow y$, and $\bar{Q}(\bm{\mu})$ for $\ket{\psi_0} = \ket{{\rm IS}_\mu}$, denoted $\mu \rightarrow \mu$) and cross probability ($\bar{Q}(\bm{y})$ for $\ket{\psi_0} = \ket{{\rm IS}_\mu}$, and viceversa, denoted $y \longleftrightarrow \mu$). While $\bar{Q}$ decays exponentially in $N$, scarring makes the return probabilities consistently larger than the cross probabilities, indeed by a factor $>4$ for all considered system sizes.
			In (a) we considered $s = \frac{1}{2}$, $N = 20$, and $\bm{\mu} = (2.4,0,0.4)$, and chose $\bm{J}$ ensuring that the system is fully thermal: $J_{zz} = -1.8$ for the Ising model, $J_{xx} = J_{yy} = -0.4$ and $J_{zz} = -1.8$ for the XXZ model, and $J_{xx} = J_{yy} = -1.4$ for the XX model. In (b,c) we considered the Ising model.}
		\label{Fig4}
	\end{figure}
	
	As first shown for quantum billiards~\cite{kaplan1999scars}, scarring can be expected when $\lambda/\omega < 1$, with $\lambda$ the Lyapunov exponent of the UPO and $T = 2\pi/\omega$ its period. That is, scars are expected when chaos, acting on a timescale $\sim \lambda^{-1}$, does not prevent a classical trajectory nearby an UPO to return to its neighborhood after one period. In Fig.~\ref{Fig1}(c) we compute $\lambda/\omega$ for some cases of interest. The IS is unstable in all the considered parameter range, with $\lambda/\omega \sim |\bm{J}|/|\bm{\mu}|$, suggesting that scarring can be enhanced by simply increasing the strength of the field $|\bm{\mu}|$. The TI state also has $\lambda/\omega < 1$, but becomes stable ($\lambda = 0$) for small $|\bm{J}|$. In the following, we consider TI and IS states only when they are unstable ($\lambda>0$), which is paramount when talking about quantum scars~\cite{kaplan1999scars}.
	
	\textit{Quantum scars} ---
	Let us go back to the quantum many-body problem $\hat{H} \ket{E_n} = E_n \ket{E_n}$. We shall here focus on the deep-quantum limit of $s = \frac{1}{2}$ (larger $s$ are considered in~\cite{SM} and yield similar results). The effective size of the Hilbert space is reduced by exploiting the symmetries of the Hamiltonian and of the TI and IS states, i.e., translation by $4n$ sites and mirror reflection, but remains exponentially large in $N$ (for a detailed discussion on symmetries, see~\cite{SM}). The eigenstates, attained by exact diagonalization, fulfill the ETH and are characterized by an extensive bipartite entanglement entropy, see Fig.~\ref{Fig2}(a,b). To look for scarring we project the eigenstates onto the classical phase space. Each point $\{\bm{s}_i\}$ of the phase space is associated to a product state
	$\ket{\{\bm{s}_i\}} = \ket{\bm{s}_1} \otimes \ket{\bm{s}_2} \otimes \dots \otimes \ket{\bm{s}_N}$,
	with $\bm{s}_i$ the orientation of the $i$-th quantum spin, $(\bm{s}_i \cdot \hs_i) \ket{\bm{s}_i} = s \ket{\bm{s}_i}$; the projection of a wavefunction $\ket{\psi}$ on the classical phase space then reads $Q = \left| \langle \{\bm{s}_i\} | \psi \rangle \right|^2$. Being increasingly accessible in quantum computers and simulators, where for $s = 1/2$ they are often called bitstring probabilities, such projections are quickly emerging as a key object of investigation in many-body quantum chaos~\cite{boixo2018characterizing,arute2019quantum,mark2024maximum,andersen2025thermalization,lami2025anticoncentration,shaw2024universal,christopoulos2024universal,claeys2025fock}.
	
	The high dimensionality of the classical phase space makes visualizing $Q$ generally complicated. Nonetheless, we are mostly interested in projections along the UPOs, which lie on two-dimensional manifolds parametrized by a polar angle $\theta$ and an azimuth $\phi$. For a few eigenstates in the middle of the spectrum, in Fig.~\ref{Fig2}(c) we show the projection $Q$ on the manifolds of TI and IS states. Such projection displays a distinctive feature of scarring, namely it reflects the underlying UPOs and peaks on some. The degree of scarring, as well as which UPOs are responsible for it, varies from eigenstate to eigenstate. Scarring is particularly remarkable for the IS manifold: not only are all the considered eigenstates in the middle of the spectrum, $E_n \approx 0$, but the whole IS manifold is, because any IS state yields $\bra{\{\bm{s}_i\}} \hat{H} \ket{\{\bm{s}_i\}} = 0$. That is, the structure in $Q$ cannot be due to some UPOs being at a special energy, which makes scarring even more surprising, in analogy with quantum billiards in which the classical orbits are all at the same energy $\frac{p^2}{2m}$~\cite{heller1984bound}.
	
	The eigenstates shown in Fig.~\ref{Fig2} are selected to showcase the structure of $Q$ in its various shapes and colors. But cherry picking is by no mean required: $Q$ reflects the underlying UPOs for the majority of the eigenstates, which we also show for the XX and XXZ models in~\cite{SM}. We also note that for the Heisenberg model ($\bm{J} = J \bm{I}$, not shown), the projection $Q$ of the eigenstates $\ket{E_n}$ is exactly constant along the IS orbits, owing to the underlying $U(1)$ symmetry $[\hat{H}, \sum_j \bm{\mu} \cdot \hs_j] = 0$. By contrast, in the models considered here the structure of the eigenstates on the UPOs does not ``piggyback'' on any symmetry.
	
	Having seen in Fig.~\ref{Fig2} the visual, qualitative features of scarring, we now turn to a quantitative analysis, showing that the eigenstates are anomalously large on the UPOs. We consider the mid-spectrum eigenstates $\ket{E_n}$ of the symmetry sector with zero momentum and left-right mirror parity $+1$, and overlap them with three types of states $\ket{\psi}$:
	Haar random states,
	phase-space states $\ket{\{\bm{s}_i\}}$,
	and IS states. To make sure these are treated on par, namely that generic phase-space states are not penalized for being less symmetric than the IS states, we renormalize all the states to have weight $1$ on the considered symmetry sector (details in~\cite{SM}). That is, we consider the rescaled projection of $\ket{E_n}$ on $\ket{\psi}$, namely $x = \mathcal{D} \frac{\left| \langle  E_n | \psi \rangle \right|^2}{\langle \psi | \hat{\mathcal{P}} \hat{\mathcal{P}}^\dagger | \psi \rangle}$, where $\mathcal{D}$ is the size of the considered symmetry sector and $\hat{\mathcal{P}}^\dagger$ the operator projecting on it. Sampling $\ket{E_n}$ and $\ket{\psi}$ yields a probability distribution for $x$, shown in Fig.~\ref{Fig3}(a). Haar random states yield the Porter-Thomas distribution $e^{-x}$~\cite{porter1956fluctuations,boixo2018characterizing}, setting a benchmark for quantum chaotic behaviour. This benchmark is closely followed by the projections of the eigenstates on generic points of the phase space $\{\bm{s}_i\}$. A deviation from the benchmark is found for projections on the IS states, yielding a fat tail in the distribution of $x$. This is remarkable: it shows that, due to scarring, the eigenstates can indeed have an anomalously large projection on the UPOs.
	
	To quantify at the same time both aspects of scarring, namely the visual features of Fig.~\ref{Fig2} and the fat tail of the projection in Fig.~\ref{Fig3}(a), we introduce a ``scarness'' parameter $S = 4 \mathcal{D} \times \underset{\rm IS}{\text{max}} \oint Q_{\psi}$, where $\oint Q$ denotes averaging of $Q$ along each UPO, and $\underset{\rm IS}{\text{max}}$ maximization over the IS UPOs (details in~\cite{SM}). Sampling $\ket{\psi}$ yields a probability distribution for $S$, shown in Fig.~\ref{Fig3}(b). While the mid-spectrum eigenstates are often compared to Haar random states, they yield a distribution of the scarness $S$ with a much fatter tail. Indeed, the $S$ of many eigenstates is larger than the $S$ of most Haar states, and the number of scarred eigenstates $M_{\rm scarred}$ grows exponentially with the considered system sizes $N$ (inset).
	
	The eigenstates are not directly measurable, but the dynamics is, and this must reflect the properties of the eigenstates. Indeed, while the thermal spectrum underpins the equilibration of local observables within short times, the scarring of exponentially many eigenstates leaves a mark on the long-time dynamics of the return probabilities. In analogy to single-particle quantum chaos~\cite{kaplan1999scars}, preparing the system on a UPO increases the probability of finding it on the same UPO at later times. This effect is shown in Fig.~\ref{Fig4} by considering two initial conditions: the IS state with axis $\bm{y}$, namely
	$\ket{\psi_0} = \otimes_{i = 1}^N \ket{\nu_i \bm{y}}$, and the IS state with axis $\bm{\mu}$, namely $\ket{\psi_0} = \otimes_{i = 1}^N \ket{\nu_i \bm{\mu}/|\bm{\mu}|}$, with $\{\nu_i\} = \left(++--++--\dots\right)$. For both we numerically integrate the Schr\"{o}dinger dynamics and compute the time-averaged projection $\bar{Q} = \lim_{t \to \infty} \frac{1}{t} \int_0^t d\tau \ Q(\ket{\psi(\tau)})$ on the manifold of IS states. For the Ising, XX, and XXZ models in a field, we observe that the system is more likely to be found on the UPO it started from, even long after thermalization.
	
	We emphasize that this effect is not due to the initial condition overlapping with a few non-thermal eigenstates~\cite{turner2018weak}, but to many of the thermal eigenstates being scarred. This is a genuinely quantum effect: due to ergodicity, a classical ensemble prepared nearby an IS UPO will at long times spread uniformly across the phase space at $E = 0$, in a way that does not depend on the specific initial condition, see Fig.~\ref{Fig4}(a). By striking contrast, it is more likely to find the quantum system on the UPO it started from. In other words, scarring makes the quantum system remember its past better than a classical system would, in a rare example of weak ergodicity breaking.
	
	We use the adjective \textit{weak} because, while the enhancement of the phase-space projection on the UPOs is large in relative terms, it is small in absolute terms, possibly scaling as $Q \sim N e^{-\mathcal{O}(N)}$, e.g., $Q_{\rm max} \sim 5 \times 10^{-4}$ in Fig.~\ref{Fig2} and $\bar{Q}_{\rm max} \sim 3 \times 10^{-5}$ in Fig.~\ref{Fig4}. A scaling analysis is presented in Fig.~\ref{Fig4}(b,c), showing that $\bar{Q}$ decays exponentially with $N$ while its relative enhancement does not, exhibiting a memory effect.
	Due to translation symmetry, the Hilbert space is divided in $N$ momentum sectors, each of size $\sim e^{\mathcal{O}(N)}/N$, and we expect the average overlap to scale as the inverse sector size, $Q \sim N e^{-\mathcal{O}(N)}$.
	Scarring should be seen as a small albeit measurable correction to the picture of thermalization in isolated quantum systems~\cite{polkovnikov2011colloquium,d2016quantum}, not contradicting paradigms such as the ETH~\cite{rigol2008thermalization}. Indeed, as long noted by Srednicki~\cite{srednicki1994chaos}, in many-body systems the effect of scarring is mostly washed away when integrating over the phase space, as effectively done when computing the expectation value of simple observables (e.g., $\langle s^z_j \rangle$).
	
	\section{Discussion}
	
	Analysing a broad family of spin chains, including \textit{any} uniform spin-$1/2$ chain with nearest-neighbor interactions, our work proves the ubiquity of quantum scarring in many-body systems. In single-particle systems, chaos—and consequently scarring—become meaningful only in the semiclassical regime, such as at sufficiently high energies in quantum billiards~\cite{heller1984bound}. By contrast, many-body systems can achieve a quantum chaotic regime through sufficiently large system sizes $N$. This distinction enables a fundamentally new type of scarring unique to many-body systems. For instance, we found that the $s \to \infty$ classical dynamics in Eq.~\eqref{eq. classical dynamics} scars the quantum system all the way down to the deep-quantum limit of $s = \frac{1}{2}$. Scarring enhances the eigenstates $\ket{E_n}$ along certain UPOs and makes a system better remember its past, curbing chaos even in fully thermal and non-integrable many-body systems.
	
	Our predictions can be straightforwardly verified in state-of-the-art quantum simulators for spin Hamiltonians~\cite{blatt2012quantum, zhang2017observation, bernien2017probing, monroe2021programmable, mi2022time}, opening new possibilities for the experimental observation of scars~\cite{wintgen1989irregular, sridhar1991experimental, stein1992experimental, fromhold1995manifestations, wilkinson1996observation}. In particular, by repeatedly preparing a product state, letting it evolve, and measuring it in a different basis, one should be able to show that the system is more likely to be found on the UPO it was prepared on (Fig.~\ref{Fig4}), which is a direct consequence of quantum scarring. This protocol is in very close analogy with what has been done in~\cite{boixo2018characterizing,arute2019quantum}, in which the many-body bitstring probabilities -- essentially the same as our projection $Q$ -- have been measured to detect genuine quantum effects. Indeed, accessing complex quantities not described by ETH, such as the bitstring probabilities, allows to search for new physics in spite of the thermalization of local observables, as we have proven for scarring. Our work opens many avenues for theoretical research, posing questions regarding the role of lattice geometry and interaction range, the fate of scarring in the limits $s\to \infty$ and $N \to \infty$, the effect of Hamiltonian terms beyond those in Eq.~\eqref{eq. H}, and possibilities with other fermionic and bosonic particles~\cite{hummel2023genuine}. Realizing that the atypical eigenstates in so-called ``many-body quantum scars'' are in fact not scars is not merely a matter of terminology~\cite{michailidis2020slow}: it opens the door to a whole new field of research, that of scars -- genuine scars~\cite{heller1984bound} -- in many-body systems, and paves the way to a better understanding of the quantum-classical correspondence.
	
	\textit{Note added}: Shortly after the completion of this work, the problem of genuine scarring in spin chains was also addressed in~\cite{ermakov2024periodic}.
	
	\textbf{Acknowledgements.}
	We thank A.~Buchleitner, C.~Castelnovo, A.~Das, I.~Ermakov, B.~Fine, I.~Kaminer, O.~Lychkovskiy, S.~I.~Mistakidis, S.~Moudgalya, A.~Nunnenkamp, N.~Rivera, and N.~Yao for insightful discussions on related work and comments on the manuscript. A.~P.~acknowledges support by Trinity College Cambridge. C.~B.~D was supported with the ITAMP grant No.~2116679.
	
	\textbf{Data Availability.}
	No datasets were generated or analysed during the current study.
	
	\textbf{Code Availability.}
	The code used for the current study is available from the corresponding author on reasonable request.
	
	\bibliography{QMBS}
	
	\clearpage
	
	\setcounter{equation}{0}
	\setcounter{figure}{0}
	\setcounter{page}{1}
	\thispagestyle{empty} 
	\makeatletter 
	\renewcommand{\figurename}{Fig.}
	\renewcommand{\thefigure}{S\arabic{figure}}
	\renewcommand{\theequation}{S\arabic{equation}}
	\setlength\parindent{10pt}
	
	\onecolumngrid
	
	\begin{center}
		{\fontsize{12}{12}\selectfont
			\textbf{Supplementary Information for\\``Genuine quantum scars in many-body spin systems"\\[5mm]}}
		{\normalsize Andrea Pizzi, Long-Hei Kwan, Bertrand Evrard, Ceren B.~Dag, and Johannes Knolle \\[1mm]}
	\end{center}
	\normalsize
	
	This Supplementary Information is devoted to technical derivations and complementary details. It is structured as follows. In Section I, we present some complementary results supporting the ubiquity of quantum scarring in many-body systems. In Section II we develop the tools needed to compute the Lyapunov exponent of the IS states, namely we perform a zero-th order Floquet expansion in a frame precessing around the external magnetic field $\bm{\mu}$. In Section III we compute the Lyapunov exponent for the IS UPOs in the limit of small $|\bm{J}/|\bm{\mu}|$. In Section IV we provide a few more details on computing classical phase-space projections.
	
	\section{I) Ubiquity of scars}
	
	We aim here to solidify the claim that scarring is ubiquitous in many-body spin chains, affecting many eigenstates across various models.
	
	\subsection{Qualitative, visual features of scars}
\begin{figure}
		\centering
		\includegraphics[width=\linewidth]{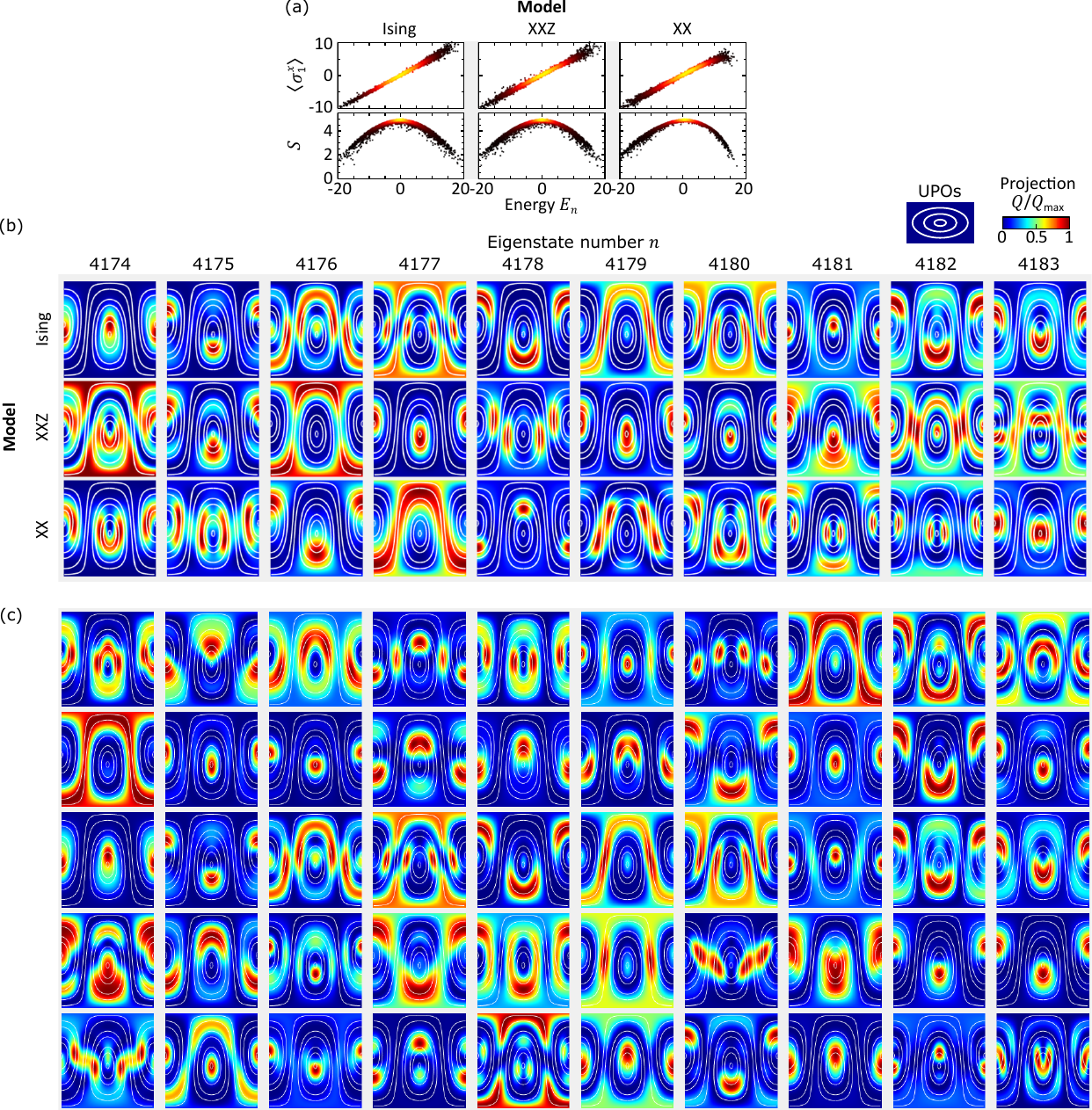}
		\caption{\textbf{Ubiquity of quantum scars.}
			(a) Expectation value $\langle E_n | \sigma_1^x | E_n \rangle$ and bipartite entanglement entropy $S$ for the eigenstates $\ket{E_n}$ of the Ising, XXZ, and XX models. (b) Projection over the manifold of IS states of 10 consecutive eigenstates taken in the middle of the spectrum. Most of the eigenstates appear scarred, to some variable extent, by the UPOs. 
			(c) Same as in (b), but for $50$ mid-spectrum eigenstates and specifically for the Ising model. The eigenstate number $n$ runs along the rows from $4154$ in the top left to $4203$ in the bottom right.
			Here, $N = 16$.}
		\label{FigS_scars}
	\end{figure}

\begin{figure}
		\centering
		\includegraphics[width=\linewidth]{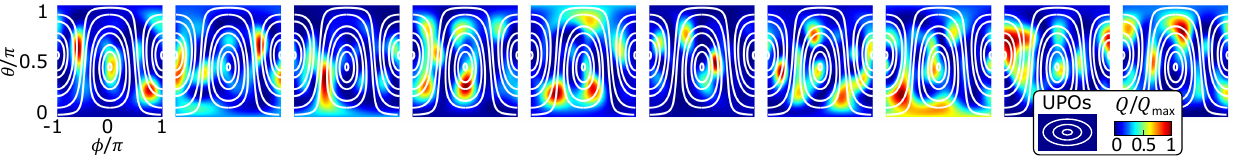}
		\caption{\textbf{Phase-space projection of Haar random states.}
			Projection $Q$ of Haar random states, for $N = 16$. The random states do not show any noteworthy structure in relation to the underlying classical UPOs.}
		\label{FigS_random}
	\end{figure}

	First, in Fig.~\ref{FigS_scars} we generalize Fig.~2 from the main text, considering the Ising, XXZ, and XX models in a field. Moreover, we do not consider a selection of eigenstates, but simply the $10$ eigenstates in the middle of the spectrum (that is composed of $8356$ eigenstates overall). The Ising model is like that in the main text, the XXZ model reads
	\begin{equation}
		\hat{H} = \frac{1}{2} \sum_{j} \left[ \bm{\mu} \cdot \hsig_j + 
		J_{xx}\left(\hat{\sigma}^x_j \hat{\sigma}^x_{j+1} + 
		\hat{\sigma}^y_j \hat{\sigma}^y_{j+1}\right) +
		J_{zz}\hat{\sigma}^z_j \hat{\sigma}^z_{j+1} \right],
		\label{eq. H XXZ}
	\end{equation}
	and the XX model reads
	\begin{equation}
		\hat{H} = \frac{1}{2} \sum_{j} \left[ \bm{\mu} \cdot \hsig_j + 
		J_{xx}\left(\hat{\sigma}^x_j \hat{\sigma}^x_{j+1} + 
		\hat{\sigma}^y_j \hat{\sigma}^y_{j+1}\right) \right].
		\label{eq. H XX}
	\end{equation}
	The specific values of the parameters are the same as in Fig.~2 in the main text, and ensure that the system is well thermalized. The expectation value $\langle E_n | \sigma_1^x | E_n \rangle$ and the bipartite entanglement entropy $S$ for the eigenstates $\ket{E_n}$ is shown in Fig.~\ref{FigS_scars}, and confirms thermalization. The projection of the 10 central eigenstates onto the manifold of IS UPOs is shown in Fig.~\ref{FigS_scars}(b). To various extents, for the majority of the eigenstates the projection $Q$ reflects the underlying UPOs. More mid-spectrum eigenstates are shown, for the Ising model, in Fig.~\ref{FigS_scars}(c). By contrast, in Fig.~\ref{FigS_random} we see that the projection $Q$ of Haar random states does not reflect in a any way the structure of the UPOs (indeed, the random states are agnostic of the Hamiltonian, and hence of the UPOs).
	
	\subsection{Quantitative features of scarring: details and role of the symmetries}
	We now turn to the question how to quantify scarring, and in particular whether the eigenstates have an anomalously large projection $Q$ on the UPOs. This is nontrivial, because the meaning of ``large'' is subtle: just out of random fluctuations, even the projection of a Haar random wavefunction happens to be somewhat larger in some points of the phase space, as in Fig.~\ref{FigS_random}, and indeed have some finite probability to be arbitrarily close to the maximum value $1$. In what sense then, exactly, can the projection of a scarred eigenstate on a UPO be said ``anomalously large''?
	
	The first and mandatory step to sensibly address these questions is to analyze the symmetries of the problem. The system is invariant under translation, which splits the Hamiltonian in sectors with a well defined momentum $k = \frac{2\pi}{N} n$, with $n = 0,1,2, \dots N-1$. Due to the left-right mirror reflection symmetry, the sectors with $k = 0$ and $k = \pi$ can be further divided into blocks with parity $P = \pm 1$. The symmetry sectors can thus be tagged by their momentum $k$ and, if $k=0$ or $k = \pi$, by a superscript $\pm$ indicating the parity. Let us call $\hat{\mathcal{P}}_\nu$ the operator that projects from the full Hilbert space to the $\nu$-th sector. We denote $\mathcal{D}_\nu$ the size of the $\nu$-th sector, and $\mathcal{D} = \sum_\nu \mathcal{D}_\nu$ the total size of the Hilbert space. For instance, for $s = 1/2$ and $N = 20$ we find $\mathcal{D}_{0^+} = 27012$, $\mathcal{D}_{\pi^+} = 25984$, $\mathcal{D}_{0^-} = 25476$, $\mathcal{D}_{\pi^-} = 26496$, $\mathcal{D}_{\pi/2} = 52380$, and $\mathcal{D} = 1048576$.
	
	It is easy to verify that the TI states are fully contained within the $0^+$ sector while the IS states distribute their weight equally within four ``special'' sectors $0^+$, $\pi^+$, $+\frac{\pi}{2}$, and $-\frac{\pi}{2}$. It follows that one can talk about scarring from the TI UPOs only for the eigenstates of the sector $0^+$, and scarring from the IS UPOs only for the eigenstates of the sectors $0^+$, $\pi^+$, $+\frac{\pi}{2}$, and $-\frac{\pi}{2}$. This is why in Fig.~2 in the main text we considered the eigenstates of the four special sectors, while in Fig.~2(c) we considered eigenstates of the sector $0^+$ (the only sector that can be scarred by the TI and IS states at the same time).
	
	These symmetry considerations done, let us go back to the question whether the projection of the eigenstates on the UPOs is ``large'', focussing on one of the four special sectors. We consider the $10\%$ most central (i.e., with smallest $|E_n|$) eigenstates of the sector. Let us call $M_e$ the number of such mid-spectrum eigenstates, for instance $M_e = 2701$ for $s = 1/2$, $N = 20$, and sector $\nu = 0^+$. As in Fig.~3 in the main text, to quantify scarring we want to compare the overlap of the mid-spectrum eigenstates with various families of states. The first are Haar random states, that  should be sampled within the symmetry sector of interest. The second are generic phase space states $\ket{\{\bm{s}_i\}}$, drawn upon sampling the spins $\{\bm{s}_i\}$ uniformly and independently from the surface of a unit sphere. The third and last are IS states, which are also of the form $\ket{\{\bm{s}_i\}}$, but for which $\bm{s}_1$ is sampled uniformly from the surface of a sphere while all the other spins are set following the IS condition, namely $\bm{s}_i = \nu_i \bm{s}_1$ with $\nu_i = (++--++--\dots)$. Crucially, generic phase-space states $\ket{\{\bm{s}_i\}}$ lack any symmetry, and their weight is thus distributed among all symmetry sectors, e.g., only $\approx 1/(2N)$-th of their weight is on the $0^+$ sector on average. By contrast, the IS states are highly symmetrical and have a relatively large weight $1/4$ on each of the special symmetry sectors. Just out of symmetry arguments it follows that an eigenstate of one of the special sectors will tend to have a larger overlap with an IS state than with a generic phase space state, on average by a factor $\sim \frac{1/4}{1/(2N)} = \frac{N}{2}$ for sectors $0^+$ and $\pi^+$, and by a factor $\sim \frac{1/4}{1/(N)} = \frac{N}{4}$ for sectors $+\frac{\pi}{2}$ and $-\frac{\pi}{2}$. Of course, such an enhancement is first of all an effect of the symmetries, not of scarring.
	
	To distill the enhancement due to scarring, we effectively cancel the effect of the symmetries by projecting out the part of the wavefunction that is not in the symmetry sector of interest. That is, given a state $\ket{\psi}$ (e.g., a phase-space state or a IS state), we define the state
	\begin{equation}
		\ket{\psi}_{\nu} = \frac{\hat{\mathcal{P}}^\dagger_\nu \ket{\psi}}{\sqrt{\bra{\psi} \hat{\mathcal{P}}_\nu \hat{\mathcal{P}}^\dagger_\nu \ket{\psi}}},
		\label{eq. projected}
	\end{equation}
	that, by construction, has weight $1$ on the symmetry sector of interest $\nu$, and $0$ on the other sectors. Thus, while phase-space states and IS states have on average a different weight on $\nu$, their projected versions in Eq.~\eqref{eq. projected} by construction have the same weight $1$, and can thus be compared fairly. Furthermore, now that all the states of interest have weight $1$ on the considered sector $\nu$, we can also conclude that their average weight on the basis states of such sector is $\mathcal{D}_\nu^{-1}$. For a sector of interest $\nu \in (0^+, \pi^+, +\frac{\pi}{2}, -\frac{\pi}{2})$, an eigenstate $\ket{E}_n$ of such sector, and a state $\ket{\psi}$,  we can thus define the rescaled overlap
	\begin{equation}
		x = \mathcal{D}_\nu \left| \langle E_n | \psi \rangle_\nu \right|^2
		= \mathcal{D}_\nu \frac{\left| \langle  E_n | \psi \rangle \right|^2}{\langle \psi | \hat{\mathcal{P}}_\nu \hat{\mathcal{P}}_\nu^\dagger | \psi \rangle},
	\end{equation}
	that is what we have used in the main text for the case of $\nu = 0^+$. Specifically, to quantify the overlap of the eigenstates with a certain family of states (Haar, phase-space states, or IS states), we generate an ensemble of rescaled projections $x$ upon sampling both $\ket{E_n}$ (from the mid-spectrum eigenstates) and $\ket{\psi}$ (from the specific family of states). For each family of states this results in a probability distribution for $x$, which we have plotted in Fig.~3.
	
	As mentioned above, scarring manifests in $Q$ having large fluctuations and having a structure in phase space that reflects the underlying UPOs. To quantify both effects at once, and focusing on the IS case, we consider the average $Q$ over an IS orbit, which we denote $\underset{\rm IS}{\oint} Q$, and look for the IS orbit that maximizes it. For a wavefunction $\ket{\psi}$ with projection $Q$ we thus introduce the following figure of merit for scarring:
	\begin{equation}
		S = 4 \mathcal{D}_\nu \times \underset{\rm IS \ orbits}{\text{max}}
		\underset{\rm IS}{\oint} Q,
	\end{equation}
	where $\mathcal{D}_\nu$ is the size of the considered special symmetry sector and the factor $4$ compensates for the fact that the IS states have weight $1/4$ on it (these normalization factors are anyway arbitrary and play no role in the following analysis). We compute $S$ for the mid-spectrum eigenstates and for an ensemble of Haar random states, which serves as a benchmark. The probability distribution of the resulting $S$ is shown in the main Fig.~3(b). While the mid-spectrum eigenstates of quantum chaotic Hamiltonians are often compared to Haar random states, and while the distributions obtained for the two indeed have some overlap, the distribution for the eigenstates is shifted towards larger values of $S$ compared to that of the Haar random states. For instance, for $s = 1/2$, $N = 20$, and sector $0^+$ we find that $\langle S \rangle_E - \langle S \rangle_H \approx 1.98 \sqrt{\langle S^2 \rangle_H - \langle S \rangle_H^2}$, where $\langle \dots \rangle_E$ and $\langle \dots \rangle_H$ denote average over the ensembles of mid-spectrum eigenstates and Haar random states, respectively. That is, the average $S$ for the eigenstates is two standard deviations larger than the average $S$ for the Haar random states.
	
	\begin{figure}
		\centering
		\includegraphics[width=\linewidth]{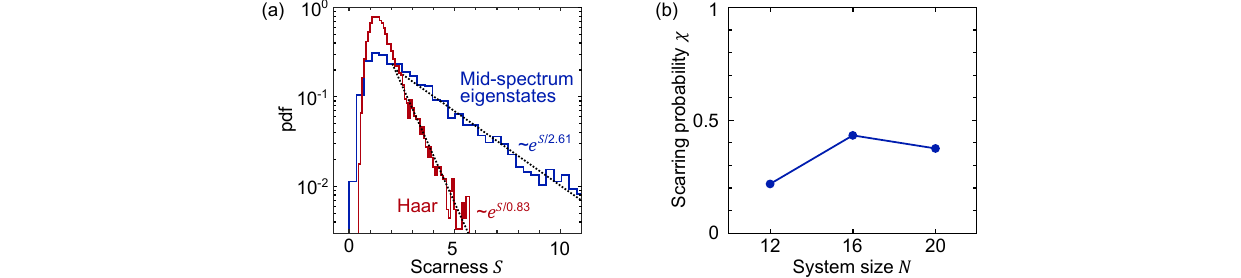}
		\caption{\textbf{Scarness distribution and probability.}
			(a) We reproduce Fig.~3(b) from the main text for the probability distribution of the scarness $S$, but with logarithmic ordinate axis. This highlights an exponential scaling at large $S$, $\sim e^{S/S_0}$. From fits (dotted) we extract $S_0 = 0.83$ for the Haar random states, and $S_0 = 2.61$ for the eigenstates. (b) Scarring probability $\chi$. While the number of accessible system sizes is limited, $\chi$ is shown to be rather large, around $40\%$ for $N = 16$ and $N = 20$, and roughly constant (or at least not strongly decaying with $N$), consistently with the observed exponential scaling of $M_{\rm scarred}$ in the inset of Fig.~3(b). Here, we consider the same parameters as in Fig.~3(b).}
		\label{FigS_scarness}
	\end{figure}
	
	This mismatch can be used to quantify the number of scarred eigenstates. To this end we consider the $M_e$ mid-spectrum eigenstates, say $M_e^\prime$ the number of them that have $S$ larger than the $90$-th percentile of the $S$ for the Haar random states, and introduce the scarring probability $\chi$ as
	\begin{equation}
		\chi = \frac{M_e^\prime}{M_e} - 0.1,
	\end{equation}
	where the $-0.1$ compensates for the fact that even for random Haar states $S$ is larger than the 90-th percentile $10\%$ of the times (by definition of $90$-th percentile). If the eigenstates behaved like Haar random states, one would get $\chi = 0$. But because, instead, many of the eigenstates have $S$ larger than most of the Haar random states, the scarring probability $\chi$ can then take relatively large values ($\sim 40\%$ for the considered parameters), indicating that a significant fraction of the eigenstates is scarred by the IS UPOs. Analogously, we can define the number of scarred mid-spectrum eigenstates as
	\begin{equation}
		M_{\rm scarred} = M_e^\prime - 10\% M_e = \chi M_e,
	\end{equation}
	which we plotted in the inset of Fig.~3(b) in the main text.
	
	Note that our analysis has focussed on the mid-spectrum eigenstates, the number of which, $M_e$, depends on how wide the central energy window is chosen, which is arbitrary. The $M_{\rm scarred}$ that we considered does not count the eigenstates outside of such energy window. Because the non-mid-spectrum eigenstates could also be scarred, one might say that $M_{\rm scarred}$ underestimates the number scarred eigenstates. At the end of the day, quantifying scarring is not an easy task, and there is no unique way of doing it. The analysis above succeeds in indicating that scarring affects many eigenstates, but for instance does not consider that certain eigenstates are scarred by multiple UPOs. There is thus room for a more systematic quantitative analysis of scarring in many-body systems, which goes however beyond the scope of this work.
	
	\subsection{Larger spins}
	
\begin{figure}
		\centering
		\includegraphics[width=\linewidth]{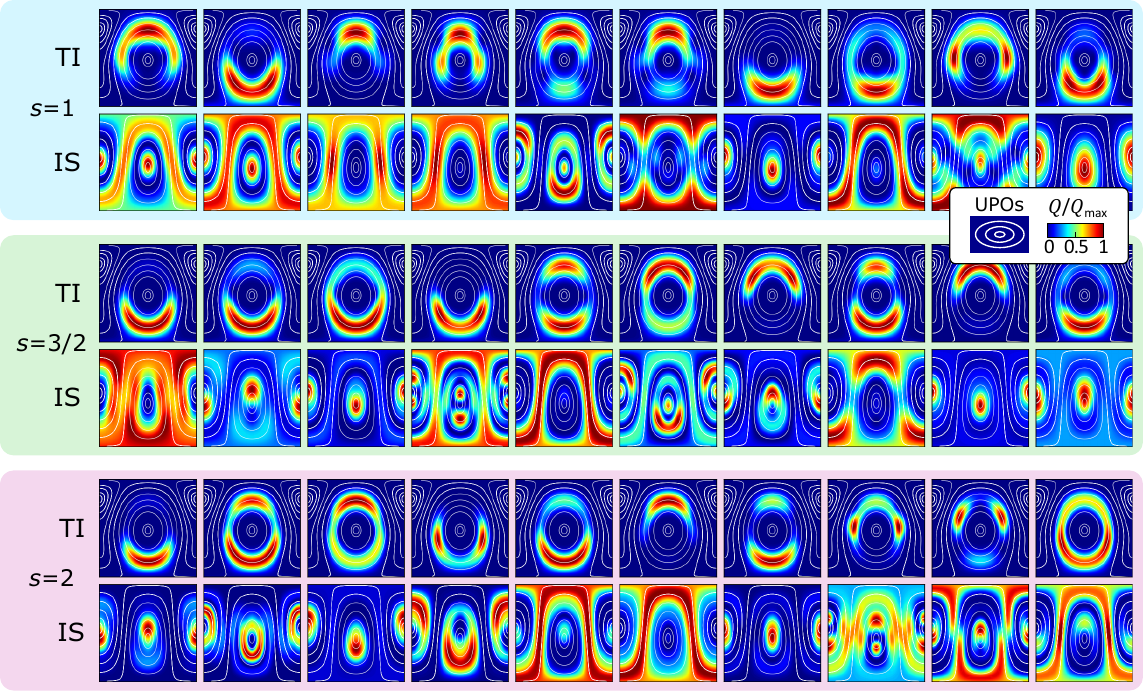}
		\caption{\textbf{Quantum scars for larger spins.}
			Projection over the manifolds of TI and IS states of 10 consecutive eigenstates taken in the middle of the $0^+$ symmetry sector for system size $N = 8$ and spin lengths $s = 1, 3/2$, and $2$. Similar to the case of $s = 1/2$, also for larger spins $s$ the eigenstates tend to reflect the structure of the underlying UPOs. Here, we considered the mixed-field Ising model with parameters as in Fig.~2 in the main text.}
		\label{FigS_scars_larger_spin}
	\end{figure}

\begin{figure}
		\centering
		\includegraphics[width=\linewidth]{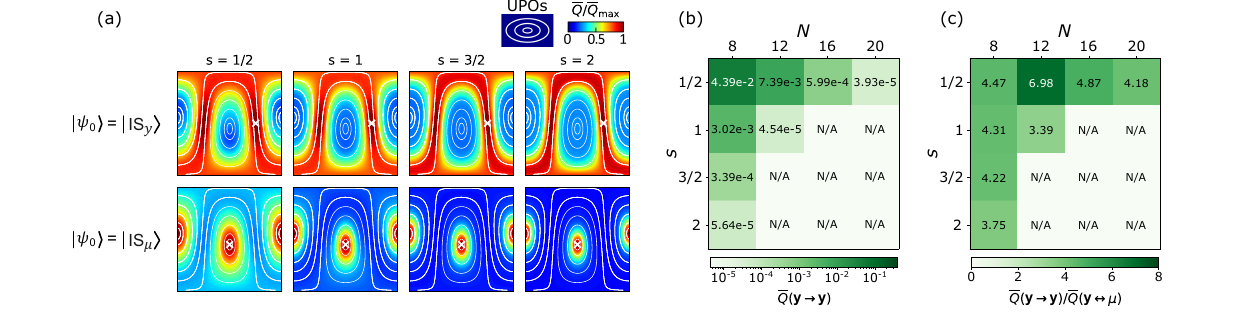}
		\caption{\textbf{Weak ergodicity breaking for larger spins.}
			(a) Time-averaged projection $\bar Q$ over the manifold of IS states for $s = 1/2$, $1$, $3/2$, $2$, and system size $N=8$. The system is initialized on the IS UPOs aligned along $\boldsymbol y$ and $\boldsymbol \mu$ in the upper and lower rows, respectively, as marked by the white cross. For all considered $s$, the time-averaged projection is enhanced on the UPO the system is initialized on, that is, ergodicity is weakly broken. The regions of larger $\bar Q$ become increasing concentrated along the UPOs when increasing $s$. (b) Scaling of the return probability $\bar Q(\boldsymbol y \rightarrow \boldsymbol y)$ with $s$ and $N$. (c) Ratio of return probability $\bar Q(\boldsymbol y \rightarrow \boldsymbol y)$ and cross probability $\bar Q(\boldsymbol y \rightarrow \boldsymbol \mu)$ for various $s$ and $N$ (namely those for which exact diagonalization is amenable -- the others are marked as N/A). The initial state is $\ket{\psi_0} = \ket{{\rm IS}_y}$ in all cases. The return probability remains around four times larger than the cross probability in all the cases we simulated. Here, we considered the mixed-field Ising model with parameters as in Fig.~2 in the main text.}
		\label{FigS_Qbar_larger_spin}
	\end{figure}

	In the main text we developed a general formalism valid for any spin length $s$, but tested it focusing on the most quantum case of $s = 1/2$. Here we show that a similar phenomenology of scarring emerges for larger spins $s = 1$, $3/2$ and $2$. In Fig.~\ref{FigS_scars_larger_spin} (in analogy with the main Fig.~2) we show that most of the mid-spectrum eigenstates have a projection $Q$ that reflects the structure of the underlying UPOs. In Fig.~\ref{FigS_Qbar_larger_spin}(a) (in analogy with the main Fig.~4), we show that the structure of the eigenstates underpins memory effects in the time-averaged projection $\bar{Q}$. Note that, going deeper into the classical limit (i.e., increasing $s$), the features of the eigenstates and of the time-averaged projection become narrower, in accordance with the decreases of the effective $\hbar$ and in analogy with the single-particle case~\cite{heller1984bound}. In Fig.~\ref{FigS_Qbar_larger_spin}(b,c) we show the scaling with $s$ and $N$ of the return probability, $\bar{Q}(\boldsymbol y \rightarrow \boldsymbol y)$, and of the ratio of return and cross probabilities, $\bar{Q}(\boldsymbol y \rightarrow \boldsymbol y)/\bar{Q}(\boldsymbol y \rightarrow \boldsymbol \mu)$. The return probability decays with both $N$ and $s$, according to the growth of the size of the Hilbert space, $\sim (2s+1)^N$. Yet, for all simulated $s$ the return probabilities remain clearly larger than the cross probabilities, by roughly four times, similar to Fig.~4(c) for spin $1/2$ in the main text.
	
	\section{II) Zero-th order Floquet expansion}
	In the frame precessing around $\bm{\mu}$ at the frequency of the IS UPOs, namely $\omega = \frac{2 \pi}{T} = |\bm{\mu}|$, the Hamiltonian reads
	\begin{equation}
		\hat{H}_{\rm rot}(t)
		=
		\frac{1}{s}
		\sum_{i = 1}^N \hs_i \bm{R}^{-1}(\omega t) \bm{J} \bm{R}(\omega t) \hs_{i+1}.
	\end{equation}
	The matrix $\bm{R}$ can be written using Rodrigues formula
	\begin{align}
		\bm{R}(\omega t) & = \bm{I} + \sin(\omega t) \bm{u}_{\cross} + [1 - \cos (\omega t)] (\bm{u}_{\cross})^2, \\
		& = \bm{u} \otimes \bm{u} + \sin(\omega t) \bm{u}_{\cross} + \cos(\omega t) (\bm{I} - \bm{u} \otimes \bm{u}),
	\end{align}
	where $\bm{u}_{\cross} \bm{a} = \bm{u} \cross \bm{a}$, $\bm{u} = \bm{\mu}/|\bm{\mu}|$, and where we used $\bm{I} + (\bm{u}_{\cross})^2 = \bm{u} \otimes \bm{u}$. The zero-th order Floquet expansion corresponds to substituting $\hat{H}_{\rm rot}(t)$ with its time average, $\hat{\bar{H}} = \frac{1}{T} \int_0^T dt \hat{H}_{\rm rot}(t)$. That is,
	\begin{equation}
		\hat{\bar{H}}
		=
		\frac{1}{s} \sum_{i = 1}^N \hs_i \bar{\bm{J}} \hs_{i+1},
	\end{equation}
	with
	\begin{align}
		\bar{\bm{J}}
		& = \frac{1}{T} \int_0^T dt \ \bm{R}^{-1}(\omega t) \bm{J} \bm{R}(\omega t), \\
		& = \int_0^1 dx
		\left[ \bm{u} \otimes \bm{u} \bm{J} \bm{u} \otimes \bm{u} 
		- \sin^2 (2 \pi x) \bm{u}_{\cross} \bm{J} \bm{u}_{\cross}
		+ \cos^2(2 \pi x) (\bm{I} - \bm{u} \otimes \bm{u}) \bm{J} (\bm{I} - \bm{u} \otimes \bm{u}) \right], \\
		& = 
		\bm{u} \otimes \bm{u} (\bm{u}\bm{J}\bm{u})
		- \frac{1}{2} \bm{u}_{\cross} \bm{J} \bm{u}_{\cross}
		+ \frac{1}{2} (\bm{I} - \bm{u} \otimes \bm{u}) \bm{J} (\bm{I} - \bm{u} \otimes \bm{u}).
	\end{align}
	Using the properties of the cross product we find
	\begin{align}
		\bm{u}_{\cross} \bm{J} \bm{u}_{\cross}
		& = 
		(\bm{I} - \bm{u} \otimes \bm{u}) (\bm{u} \bm{J} \bm{u} - \Tr{\bm{J}})
		+ (\bm{I} - \bm{u} \otimes \bm{u}) \bm{J}
		(\bm{I} - \bm{u} \otimes \bm{u}),
	\end{align}
	and thus
	\begin{align}
		\bar{\bm{J}}
		& =
		\bm{u} \otimes \bm{u} (\bm{u}\bm{J}\bm{u})
		+ \frac{1}{2} (\bm{u} \otimes \bm{u} - \bm{I}) (\bm{u} \bm{J} \bm{u} - \Tr{\bm{J}}),
		\\
		& =
		\bm{u} \otimes \bm{u} \frac{1}{2}(3\bm{u}\bm{J}\bm{u} - \Tr{\bm{J}})
		- \bm{I} \frac{1}{2} (\bm{u} \bm{J} \bm{u} - \Tr{\bm{J}}).
	\end{align}
	
	It will prove useful to further characterize the matrix $\bar{\bm{J}}$. Its eigenvectors are $\bm{u}$ and any two vectors orthogonal to it, with eigenvalues
	\begin{equation}
		\lambda_1 = 
		\bm{u}\bm{J}\bm{u},
		\quad \quad
		\lambda_2 = \lambda_3 =
		- \frac{1}{2} (\bm{u} \bm{J} \bm{u} - \Tr{\bm{J}}),
	\end{equation}
	and so
	\begin{align}
		\det(\bar{\bm{J}}) &=
		\frac{1}{4}(\bm{u}\bm{J}\bm{u})(\bm{u} \bm{J} \bm{u} - \Tr{\bm{J}})^2, \\
		\Tr{\bar{\bm{J}}} &=
		\Tr{\bm{J}}.
	\end{align}
	Moreover, we note that 
	\begin{equation}
		\left[  \bm{u} \otimes \bm{u} (3\bm{u}\bm{J}\bm{u} - \Tr{\bm{J}})
		- 2 (\bm{u} \bm{J} \bm{u}) \bm{I} \right]
		\bar{\bm{J}}
		=  (\bm{u} \bm{J} \bm{u}) (\bm{u} \bm{J} \bm{u} - \Tr{\bm{J}}) \bm{I},
	\end{equation}
	and thus find the adjoint ${\rm Adj}(\bar{\bm{J}})$ such that $({\rm Adj}(\bar{\bm{J}}) ) \bar{\bm{J}} = \det(\bar{\bm{J}}) \bm{I}$, namely
	\begin{equation}
		{\rm Adj}(\bar{\bm{J}})
		= \frac{1}{4} (\bm{u} \bm{J} \bm{u} - \Tr{\bm{J}})
		\left[  \bm{u} \otimes \bm{u} (3\bm{u}\bm{J}\bm{u} - \Tr{\bm{J}})
		- 2 (\bm{u} \bm{J} \bm{u}) \bm{I} \right].
		\label{eq. adj}
	\end{equation}

	\section{III) Lyapunov exponent}
	We compute the classical Lyapunov exponent associated to the IS UPOs, in which the spin direction flips every other spin, $\{\bm{s}_j\} = (+\bm{s},+\bm{s},-\bm{s},-\bm{s},+\bm{s},+\bm{s},-\bm{s},-\bm{s},\dots)$. Various values of $\bm{s}$ correspond to various UPOs, and we should thus use $\bm{s}$ as a parameter, a tag of the considered UPO. Computing the Lyapunov exponent with respect to a periodic orbit can be complicated, but by moving to the frame precessing around the magnetic field $\bm{\mu}$, and assuming that $|\bm{\mu}|/|\bm{J}| \gg 1$, we transform the UPO into a fixed point of the dynamics, which considerably simplifies the computation of the Lyapunov exponent. In such frame the classical spin dynamics reads
	\begin{equation}
		\frac{d \bm{s}_i}{dt} = \left(\bar{\bm{J}}(\bm{s}_{i-1} + \bm{s}_{i+1}) \right) \cross \bm{s}_i.
	\end{equation}
	We want to study the evolution of small perturbation on top of the UPO. Let us say $\bm{s}_i = \nu_i \bm{s} + \bm{\epsilon}_i$, with $\{\nu_i \} = (++--++--\dots)$ and $\bm{\epsilon}_i$ a small perturbation of order $\epsilon \ll 1$. Because $\nu_{i-1} + \nu_{i+1} = 0$, then $|\bm{s}_{i-1} + \bm{s}_{i+1}| \sim \epsilon$, and thus at first order in $\epsilon$ we can consider
	\begin{align}
		\frac{d \bm{\epsilon}_i}{dt}
		& \approx - \nu_i \bm{s} \cross [ \bar{\bm{J}} (\bm{\epsilon}_{i-1} + \bm{\epsilon}_{i+1}) ], \\
		& = \nu_i \bm{M} (\bm{\epsilon}_{i-1} + \bm{\epsilon}_{i+1}) ],
	\end{align}
	where $\bm{M} = - \bm{s}_{\times} \bar{\bm{J}}$. We perform a Fourier transformation
	\begin{equation}
		f_k = \sum_j e^{-i k j} f_j,
		\quad \quad \quad
		f_j = \frac{1}{N} \sum_k e^{i k j} f_k,
	\end{equation}
	and, noting that
	$\nu_j = - \frac{1 + i}{2} e^{i \pi/2 j} - \frac{1 - i}{2}e^{- i \pi/2 j}$,
	get
	\begin{align}
		(\nu_j (\bm{\epsilon}_{j-1} + \bm{\epsilon}_{j+1}))_k
		&= \sum_j e^{-ikj} \nu_j (\bm{\epsilon}_{j-1} + \bm{\epsilon}_{j+1}), \\
		&= - \sum_j e^{-ikj} \left( \frac{1 + i}{2} e^{i \pi/2 j} + \frac{1 - i}{2}e^{- i \pi/2 j}\right) (\bm{\epsilon}_{j-1} + \bm{\epsilon}_{j+1}), \\
		&= - \sum_j \left( \frac{1 + i}{2} e^{-i(k-\pi/2)j} + \frac{1 - i}{2} e^{-i(k+\pi/2)j}\right) (\bm{\epsilon}_{j-1} + \bm{\epsilon}_{j+1}), \\
		&= - (1+i) \cos(k - \frac{\pi}{2}) \bm{\epsilon}_{k-\pi/2} - (1-i) \cos(k + \frac{\pi}{2}) \bm{\epsilon}_{k+\pi/2},
		\\
		&= - (1+i) \sin(k) \bm{\epsilon}_{k-\pi/2} + (1-i) \sin(k) \bm{\epsilon}_{k+\pi/2}.
	\end{align}
	The eigenproblem reads
	\begin{equation}
		\frac{d \bm{\epsilon}_k}{dt}
		=
		\lambda \bm{\epsilon}_k
		=
		- (1+i) \sin(k) \bm{M} \bm{\epsilon}_{k-\pi/2} + (1-i) \sin(k) \bm{M} \bm{\epsilon}_{k+\pi/2},
		\label{eq. k-eigenproblem}
	\end{equation}
	that is,
	\begin{equation}
		\lambda
		\begin{pmatrix}
			\bm{\epsilon}_k \\
			\bm{\epsilon}_{k-\pi/2} \\
			\bm{\epsilon}_{k+\pi/2} \\
			\bm{\epsilon}_{k+\pi}
		\end{pmatrix}
		=
		\begin{pmatrix}
			0 & - (1+i) \sin(k) \bm{M} & (1-i) \sin(k) \bm{M} & 0 \\
			(1-i) \cos(k) \bm{M} & 0 & 0 & -(1+i) \cos(k) \bm{M} \\
			(1+i) \cos(k) \bm{M} & 0 & 0 & (i-1) \cos(k) \bm{M} \\
			0 & (i-1) \sin(k) \bm{M} & (1+i) \sin(k) \bm{M} & 0
		\end{pmatrix}
		\begin{pmatrix}
			\bm{\epsilon}_k \\
			\bm{\epsilon}_{k-\pi/2} \\
			\bm{\epsilon}_{k+\pi/2} \\
			\bm{\epsilon}_{k+\pi}
		\end{pmatrix}.
		\label{eq. 12-dim eigenproblem}
	\end{equation}
	That is, we have obtained $N/4$ sets of 12-dimensional eigenproblems, for a total of $3N$ eigenvalues. Each eigenproblem is associated to a set of four momenta, namely $k, k-\frac{\pi}{2}, k + \frac{\pi}{2},$ and $ k + \pi$. To label each set we can consider values of $k$ up to $\frac{\pi}{2}$, namely $k = \frac{2 \pi}{N} \times \left(1,2,\dots,\frac{N}{4}\right)$, to which we will henceforth restrict. With the assistance of a computer, we find that the $12$ eigenvalues of Eq.~\eqref{eq. 12-dim eigenproblem} are
	$\left(0,0,0,0, a e^{+i \frac{\pi}{4}}, a e^{+i \frac{\pi}{4}}, a e^{-i \frac{\pi}{4}}, a e^{-i \frac{\pi}{4}}, a e^{+i \frac{3\pi}{4}}, a e^{+i \frac{3\pi}{4}}, a e^{-i \frac{3\pi}{4}}, a e^{-i \frac{3\pi}{4}}\right)$, where $a$ is some real positive number. To find it, we iterate Eq.~\eqref{eq. k-eigenproblem} once to get
	\begin{align}
		\lambda^2 \bm{\epsilon}_k
		& =
		- (1+i) \sin(k) \cos(k) \bm{M}^2
		\left( - (1+i) \bm{\epsilon}_{k-\pi} + (1-i) \bm{\epsilon}_{k} \right)
		- (1-i) \sin(k)\cos(k) \bm{M}^2
		\left( - (1+i) \bm{\epsilon}_{k} + (1-i) \bm{\epsilon}_{k+\pi} \right), \\
		& =
		2i \bm{M}^2 \sin(2k)
		\bm{\epsilon}_{k+\pi},
	\end{align}
	which we iterate again, to close the equation, getting
	\begin{equation}
		\lambda^4 \bm{\epsilon}_k = -4 \bm{M}^4 \sin^2(2k) \bm{\epsilon}_k.
	\end{equation}
	Again with the help of a computer, we find that $\bm{M}^2$ has one vanishing eigenvalue and two degenerate eigenvalues $\alpha = - \bm{s} [\text{adj}(\bar{\bm{J}})] \bm{s}$, thus getting $a = \sqrt{2 |\alpha \sin(2k)|}$.
	
	Putting all the results together, we thus find that the $3N$ eigenvalues of the problem are $0$, with multiplicity $N$, and 
	\begin{equation}
		\lambda_{k,m} = \sqrt{2 |\alpha \sin(2k)|} e^{i \frac{\pi}{4} + i m\frac{\pi}{2}},
	\end{equation}
	with multiplicity $2$ and for $m = 1,2,3,4$ and for $k = \frac{2 \pi}{N} \times \left(1,2,\dots,\frac{N}{4}\right)$. Note: the fact that $N$ of the $3N$ eigenvalues vanish is simply due to the fact that the three components of the spins are not independent, but constrained by $\left| \bm{s}_i \right|^2 = 1$. Note as well that for $N = 4$ all the Lyapunov exponents vanish, $\lambda_{k,m} = 0$, consistently with Ref.~\cite{steinigeweg2009heisenberg}. Here, we should focus on $N \ge 8$.
	
	The instability exponent $\lambda$ is the largest of the real parts of the eigenvalues, obtained for $k = \frac{\pi}{4}$ and giving $\lambda^2 = |\alpha| = |\bm{s} [\text{adj}(\bar{\bm{J}})] \bm{s}|$. Using the expression found in Eq.~\eqref{eq. adj} we get
	\begin{align}
		\lambda^2
		& = \frac{1}{4} \left| (\bm{u}\bm{J}\bm{u} - \Tr{\bm{J}})
		\left[  (\bm{u} \cdot \bm{s})^2 (3\bm{u}\bm{J}\bm{u} - \Tr{\bm{J}})
		- 2 \bm{u}\bm{J}\bm{u} \right] \right|.
		\label{Eq. S_lambda}
	\end{align}
	For instance, for the Ising model in a field we get
	\begin{equation}
		\lambda^2
		= \frac{1}{4} J^2 u_x^2
		\left[ (1- 3u_z^2)(\bm{u} \cdot \bm{s})^2
		+ 2 u_z^2 \right].
	\end{equation}
	Note, the fact that $\lambda$ in Eq.~\eqref{Eq. S_lambda} does not depend on $N$ is not obvious a priori: it is a special feature of the considered IS UPOs, for which the dynamics decouples for groups of just four momenta in Eq.~\eqref{eq. 12-dim eigenproblem}.
	
	\section{IV) Classical fidelity}
	The overlap of a classical state $\ket{\{\bm{s}_j^\prime \}}$ onto a point $\{\bm{s}_j\}$ of the classical phase space reads
	\begin{equation}
		Q = \left|\langle  \{\bm{s}_j \} | \{\bm{s}^\prime_j \} \rangle \right|^2
		= \prod_j \left|\langle  \bm{s}_j | \bm{s}^\prime_j \rangle \right|^2
		= \prod_j \left( \frac{1 + \bm{s}^\prime_j \cdot \bm{s}_j}{2} \right)^{2s}.
	\end{equation}
	We note in passing that $Q \approx \exp(- \frac{s}{2} \sum_j \theta_j^2)$, with $\theta_j$ the angle between $\bm{s}_j$ and $\bm{s}_j^\prime$ and where the approximation becomes more accurate when increasing $s$. Indeed, the states $\ket{\{\bm{s}_j\}}$ are the natural generalization of Gaussian states for spins.
	
	Consider a point $\{\bm{s}_j\}$ of the classical phase space. A perturbation of this point is obtained by rotating each spin by an angle $\theta_j$ drawn at random from a Gaussian distribution with standard deviation $\Delta  = 10^{-5}$ (almost identical results are obtained for any $\Delta \ll 1$, provided that the time averaging is performed over a time $\gg \frac{1}{\lambda} \log \Delta$). Considering many such perturbations we obtain an ensemble of phase-space points $\{\bm{s}_j^{(r)}\}$, with $r = 1,2,\dots,R$, that are closely localized around the reference point $\{\bm{s}_j\}$. By classically time evolving the ensemble, we obtain the corresponding classical time-averaged projection
	\begin{equation}
		\bar{Q}_c(\{\bm{s}_j\}) = \frac{1}{R} \sum_{r = 1}^R \lim_{t \to \infty} \frac{1}{t} \int_{0}^{t} d\tau \
		\prod_j \left( \frac{1 + \bm{s}_j \cdot \bm{s}^{(r)}_j(t)}{2} \right)^{2s}.
	\end{equation}
	Note that $\bar{Q}_c(\{\bm{s}_j\})$ is exponentially small in both $N$ and $s$. The parameter $s$ controls how heavily a misalignment between $\bm{s}^{(r)}_j$ and $\bm{s}_j$ is penalized. The larger $s$, the more unlikely that a trajectory will meaningfully contribute to the sum, and the larger the number of samples $R$ required for convergence. In Fig.~4 we consider $s = 1/2$, for which a good convergence is obtained for $R = 2 \times 10^5$ samples. This choice is ultimately arbitrary, and we expect that the same key finding (that $\bar{Q}_c$ does not depend on the initial condition) would be found for larger values of $s$, although at the cost of a larger computational burden.
\end{document}